\documentclass[preprint,showpacs,showkeys,aps,prb]{revtex4}
\usepackage[dvips]{graphicx}

\begin{document}

\title{
Nonequilibrium Systems : Hard Disks and Harmonic\\
Oscillators Near and Far From Equilibrium       \\
}

\author{
William Graham Hoover and                       \\
Carol Griswold Hoover                           \\
Ruby Valley Research Institute                  \\
Highway Contract 60, Box 601                    \\
Ruby Valley, Nevada 89833 ;                     \\
Julien Clinton Sprott                           \\
Department of Physics                           \\ 
University of Wisconsin - Madison               \\
Wisconsin 53706 .                               \\
}
\date{\today}

\keywords{Chaos, Ergodicity, Thermostats, Reversibility}

\vspace{0.1cm}

\begin{abstract}
We relate progress in statistical mechanics, both at and far from
equilibrium, to advances in the theory of dynamical systems.
We consider computer simulations of time-reversible deterministic
chaos in {\it small} systems with three- and four-dimensional phase
spaces.  These models provide us with a basis for understanding
equilibration and thermodynamic irreversibility in terms of Lyapunov
instability, fractal distributions, and thermal constraints.
\end{abstract}

\maketitle

\section{Introduction}
Nonequilibrium Molecular Dynamics and Dynamical Systems Theory have been our
main research interests for about 50 years, the same period over which Moore's
Law has described the growth of our primary tool, computation.  In 1959
thermodynamic information was mainly gleaned from series expansions of
pressure in powers of the density and integral equations for the pair
distribution function. That was the year when Berni Alder and Tom Wainwright
described a new simulation method\cite{b1} now called ``molecular dynamics'' in
their prescient Scientific American article ``Molecular Motions'' :
\begin{quote}
``One of the aims of molecular physics is to account for the bulk properties
of matter [ pressure $P$, temperature $T$, energy $E$, \dots \ ] in terms of
the behavior of its
particles. High-speed computers are helping physicists realize this goal.''
\end{quote}

At that time simulating the motion of a few hundred particles presented a
computational challenge.  Today's biomolecule simulations model at least many
thousand and perhaps a few million atomistic degrees of freedom. After several
Nobel prizes\cite{b2} this molecular dynamics method is familiar textbook material
while the virial series for the pressure and the pair-distribution integral
equations keep company with the dinosaurs.

During this same period our understanding of dynamical systems ( flows described
by a few nonlinear ordinary differential equations ) has undergone explosive
growth.  Ed Lorenz' three-equation Butterfly Attractor is a clearcut demonstration
of ``chaos'', the exponential ``Lyapunov instability'' often found in systems of
three or more ordinary differential equations.  The Lyapunov spectrum of exponential
growth and decay rates provides a topological description of evolving phase-space
densities. The discovery that {\it time-reversible} flow equations can describe
{\it irreversibility} through the formation of fractal strange attractors furnished
a new geometric interpretation of the Second Law of Thermodynamics in terms of an
underlying reversible mechanics\cite{b3}.

The correspondence between manybody molecular dynamics and the concepts developed
in dynamical systems theory involves five key ideas :

[ 1 ] Simulating nonequilibrium systems requires a new {\it nonequilibrium molecular
dynamics} which, unlike Hamiltonian mechanics,  includes thermodynamic control variables.

[ 2 ] These control variables, such as thermostats or ergostats, can provide ergodic
equilibrium dynamics, replicating Gibbs' canonical distribution.

[ 3 ] Away from equilibrium this same approach, while time-reversible, can promote
and maintain nonequilibrium steady states.

[ 4 ] Despite the time-reversible nature of the nonequilibrium flow equations the
resulting phase-space description is {\it dissipative}, on average, and generates
multifractal attractors.

[ 5 ] The multifractal nature of nonequilibrium steady states confirms their rarity
and provides a mechanical explanation of the Second Law of Thermodynamics.

The dynamical systems approach to irreversible processes continues to provide new
insight into both equilibrium and far-from-equilibrium flows\cite{b4,b5}.  Our intent
here is to illustrate this insight by the exploration of the simplest possible
dynamical models for nonequilibrium steady states.  We begin with the Galton Board
problem\cite{b5,b6,b7}, a steady field-driven flow with impulsive hard-disk collisions.
We continue, and then conclude, with a variety of generalized harmonic oscillator
problems\cite{b8}.  These illustrate heat flow and ergodic fractal formation with just
three ordinary differential equations. The hard-disk Galton Board and the generalized
conducting-oscillator problems display all of the key ideas linking manybody
mechanics to small-system analogs.

The plan of this work is as follows.  We first review the Galton Board problem and
use that example to illustrate the fractal attractors generated by time-reversible
nonequilibrium steady states.  We visualize these attractors through two-dimensional
cross sections of their three-dimensional phase-space distributions.  The Galton
Board is one of the simplest chaotic problems.  It is deterministic and ergodic in
its three-dimensional phase space.  The ergodicity is enabled by the ( exponential )
Lyapunov instability of its hard-disk collisions.

We then explore ergodicity ( dynamical access to {\it all} phase-space states ) for
smoothly-continuous harmonic-oscillator problems, at and away from thermal equilibrium.
Many of the nonequilibrium versions of oscillator problems provide dissipative strange
attractors in just three or four phase-space dimensions.  We point out some useful
numerical techniques for exploring the boundary between chaos and regularity and
discuss the possibility of numerical implementations of Liouville's phase-space
flow equations.  Finally we tie together these simple microscopic example problems to
their real-world analogs in macroscopic thermodynamics and computational fluid mechanics.

\section{The Galton Board--Ergodic, Time-Reversible, Dissipative}

\begin{figure}
\includegraphics[width=4.5in,angle=90.]{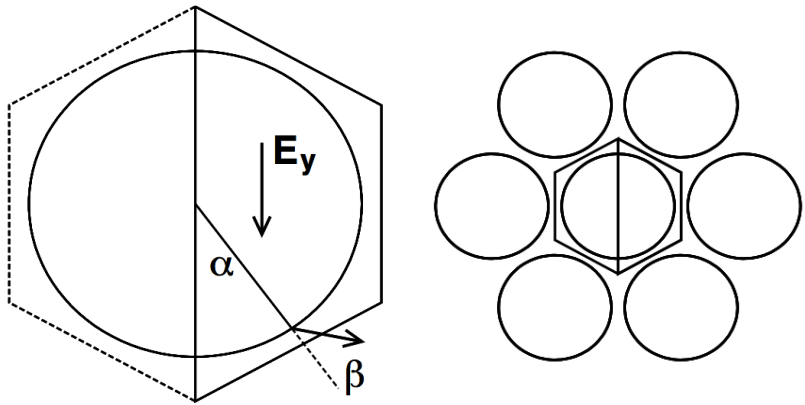}
\caption{
Description of collisions in the Galton Board in terms of the angles $\alpha$ and $\beta$ .
At a collision the radial velocity, $-\cos(\beta)$ changes sign while the tangential velocity
$\sin(\beta)$ is unchanged.
}
\end{figure}

\begin{figure}
\includegraphics[width=3.5in,angle=90.]{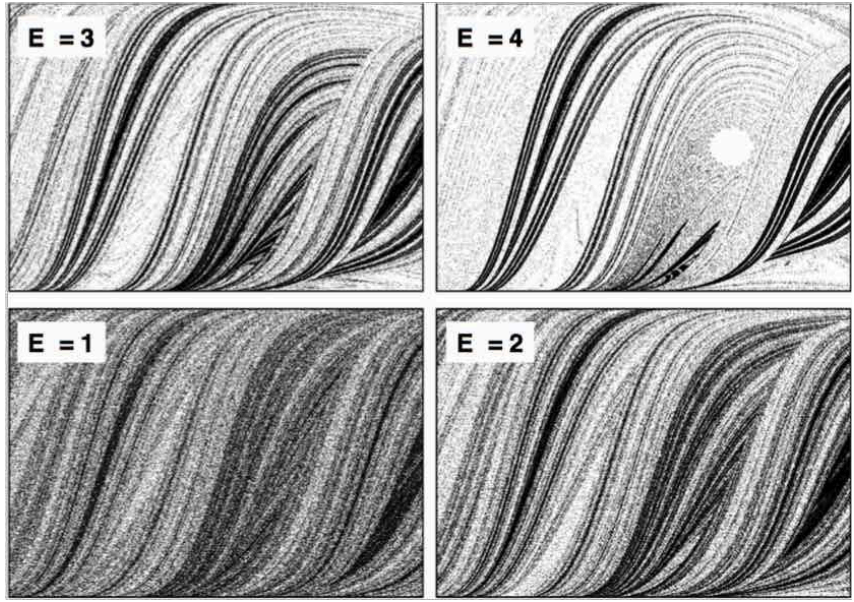}
\caption{
Field Dependence of Galton Board Collisions.  Each collision corresponds to a
single point $[ \ 0 < \alpha < \pi \ {\rm with} \  -1 < \sin(\beta) < + 1 \ ]$
so that each of these four sample phase-space cross sections illustrates 300,000
successive collisions.  At zero field strength the distribution of points is
completely uniform with a constant density of points.
}
\end{figure}

Our goal throughout is to connect dynamics, statistical mechanics, and nonequilibrium
systems with the simplest possible examples.  The Galton Board models Sir Francis
Galton's lecture table probability demonstration based on the chaotic motion of
particles introduced at the top and in the center of a fixed lattice of
scatterers. In our idealized mechanical steady-state model this field-driven motion
occurs at constant speed. To implement this idea the field's acceleration is moderated
by a deterministic time-reversible ``thermostat'' force acting parallel to the
particle's velocity.  The resulting trajectory is isokinetic and continuously
dissipates field energy as heat. The field is the source of energy. The heat reservoir
represented by a thermostat force is the compensating heat sink. The overall description
of this isothermal ( constant kinetic energy ) Galton Board model system is {\it the}
prototypical simplest mechanical example of a time-reversible nonequilibrium steady state.
The computation of its dynamics involves solving four coupled ordinary differential
equations for the $(x,y)$ location of the falling particle and its velocity $(p_x,p_y)$
and is punctuated by hard-disk scatterer collisions.  For simplicity the falling particle
has unit mass and speed. We choose the accelerating field parallel to the $y$ axis, which
is perpendicular to one set of rows of scatterer particles, as shown in {\bf Figure 1}.

The resulting diffusive motion through such a periodic array of scatterers is easy to
program, particularly if the scatterers are motionless ``hard disks''.  One simply
integrates the motion equations for $(x,y,p_x,p_y)$ until the moving particle finds
itself ``inside'' a scatterer
$( \ | \ r \ | < 1/2 \ )$ . Then the dynamics is returned to the previous $( \ x,y \ )$
coordinate set.  There the sign of the radial velocity is changed from negative to
positive, and the integration is continued.  We have found that Runge-Kutta integration
is the simplest useful method for generating trajectories between collisions.  The
alternative analytic approach\cite{b6}, though feasible, is cumbersome.

In between collisions, the equations of motion, with the field $-E$ in the $y$ direction,
are :
$$
\dot x = p_x \ ; \ \dot y = p_y \ ; \ \dot p_x = -\zeta p_x \ ; \ 
\dot p_y = -E - \zeta p_y \ .
$$
We choose the linear ``frictional force'' $-\zeta p$, to
enforce an isokinetic constant-speed constraint.  This linear force is the simplest
choice.  It also follows directly from Gauss' Principle of Least Constraint\cite{b9},
where the constraint condition controls the kinetic energy, for which we choose
$$
K \equiv (p_x^2 + p_y^2)/2 = (1/2) \ ; \ \dot K \equiv 0 \ .
$$
The linear constraint force, $-\zeta p$ , is sufficient to satisfy the isokinetic
condition :
$$
\dot K \equiv 0 = p_x\dot p_x + p_y\dot p_y = -\zeta p_x^2 -Ep_y -\zeta p_y^2
\longrightarrow \zeta = -Ep_y \ .
$$

A unit cell, within which the motion occurs, is pictured in {\bf Figure 1}.  It is
convenient to think of the moving particle as a mass point and the scatterers as
fixed particles of unit diameter.  At each collision ( defined by the two angles
$\alpha$ and $\beta$ shown in {\bf Figure 1} ), the radial component of the
velocity $[ \ -\cos(\beta) \ ]$ is reversed from negative to positive and the motion
is continued with the resulting post-collision values of $\{ \ p_x,p_y \ \}$ .  By
choosing a scatterer density of four fifths the maximum close-packed density we avoid the
possibility of a ballistic collisionless trajectory.  The inevitable scatterer collisions
make possible ( and for moderate field strengths, inevitable ) diffusive
piecewise-continuous trajectories punctuated by a series of scatterer collisions.

Three distinct types of solution result, conservative, dissipative, and periodic,
with the type determined by the initial condition and the field strength $E_y$ .
For zero field the motion is ergodic and conservative, obeying the equilibrium
version of Liouville's Theorem, $\dot f(x,y,p_x,p_y) \equiv 0$ .  That is, all
conceivable collision types do occur, and with a uniform probability when plotted
in the $[ \ \alpha,\sin(\beta) \ ]$ plane.  As the field is increased it becomes
apparent that the distribution of collisions, though becoming nonuniform, remains
``ergodic'', with nonzero probability everywhere.  We use the word ``ergodic'' as
equivalent to the Ehrenfests' ``quasiergodic'' notion of a dynamics that eventually
comes arbitrarily close to every point in the distribution.  At relatively high
values of the accelerating field things can be different.   Trajectories can become
trapped in stable periodic orbits, some conservative and some dissipative so that
the motion is no longer ergodic.  See, for an obvious example, the ``hole'' in 
the distribution for $E=4$ shown in {\bf Figure 2} .

Quantitative investigation of the two-dimensional phase-space cross sections
illustrated in {\bf Figure 2} reveals that the densities of points in the vicinity of
( 1 ) a randomly chosen  point $\propto r^{D_1}$ and of ( 2 ) a collisional point
$\propto r^{D_2}$ are different, with $2 \ge D_1 \ge D_2 \ge $ 1 . These power laws
define\cite{b4} ( 1 ) the ``information dimension'' and ( 2 ) the ``correlation dimension''
of the various fractal cross sections. Both of these two fractional dimensions vary
with field strength. 

Despite this singular fractal behavior there is no problem computing the mean vertical
current and the equivalent conductivity.  New phenomena appear at field strengths
somewhat higher than $E = 3$ : stable sequences of repeated collisions begin to occur.
In the full three-dimensional phase space, which includes an additional time dimension
for the free-flight portion of the trajectories, the stable sequences are described by
regular conservative tori in $[ \ \alpha,\sin(\beta),t \ ]$ space or by dissipative
limit cycles in which field energy is absorbed by the time-reversible friction force
$-\zeta p$ .  Examples of both types are shown in {\bf Figure 3}.  The family of
two-bounce horizontal orbits shown at the left has no net current.  In contrast, the
ten-bounce orbit on the right is strongly dissipative with a net downward current.
This motion generates  a one-dimensional {\it limit-cycle} orbit in the three-dimensional
phase space and would be represented by five {\it zero}-dimensional dots in a collisional
cross-section picture of the type shown in {\bf Figure 2} .

\begin{figure}
\includegraphics[width=5.5in,angle=90.]{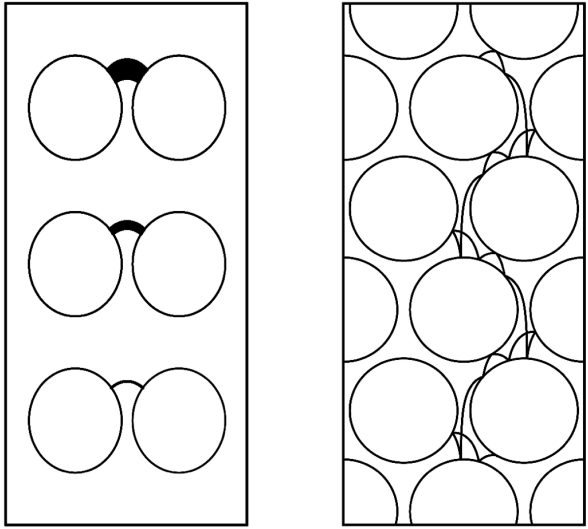}
\caption{
Conservative Tori at $E=4$ and a Dissipative Limit Cycle at $E=6$ . The two-bounce
trajectories shown at the left occur in the prominent ``hole'' seen in the $E=4$
collision plot of {\bf Figure 2} .  The periodic orbit at the right corresponds to five
zero-area points in the $[ \ \alpha,\sin(\beta) \ ]$ representation of {\bf Figure 2} .
}
\end{figure}

Let us summarize our findings from this simple nonequilibrium steady-state problem.
The Galton Board is deterministic, time-reversible, and dissipative.  With the field
``off'' the motion is ergodic -- it comes arbitrarily close to any collision type from
headon $[ \ \sin(\beta) = 0 \ ]$ to glancing  $[ \ \sin(\beta) = \pm \pi \ ]$ and
anywhere from the top $( \ \alpha = \pi \ )$ to the bottom $( \ \alpha = 0 \ )$ of a
scatterer.  This ergodicity provides a direct connection between Newton's dynamical and
Gibbs' statistical treatments of mechanical systems.

With the field ``on'' the dynamics becomes fractal, though still ergodic for moderate
field strengths.  The phase-space description becomes a fractional-dimensional
representation of collisions which is singular, nonuniform, and dissipative.  The
dissipation reflects the  conversion of gravitational work into heat through the
mechanism of a heat reservoir.  Although the motion and the motion equations are
perfectly time-reversible, the typical field-driven case is at the same time
dissipative.  The conversion of field energy $Ey$ to extracted heat is imposed by the
friction coefficient $\zeta$ :
$$
\langle \ \zeta p^2 \ \rangle  = \langle \ 2\zeta K \ \rangle = \langle \ \zeta \ \rangle
= \langle \ \dot S/k \ \rangle > 0 \ .
$$
Dividing the dissipated heat $\zeta p^2$ by the temperature, $T = (p^2/k)$ shows that
the friction coefficient is also equal to the instantaneous irreversible entropy
production, $\zeta = (\dot S/k)$ where $k$ is Boltzmann's constant.  In our numerical
work we set it equal to unity, $k \equiv 1$ .

Because the Galton Board distributions are fractal, with zero-area cross sections
[ having fractal dimensionalities less than two ] the random-sampling probability
of finding a point $[ \ \alpha,\sin(\beta) \ ]$ on the strange attractor is precisely
zero.  If one attempts to define a limiting probability density in the cross sections
by counting points in small cells, he soon discovers that the density does not have a
small-cell limiting value.  Instead it diverges as a fractional power of the cell
size.  This finding was both surprising and illuminating in 1987\cite{b3,b6}.  Since
then it has turned out that such fractal attractors are {\it typical} representations of
nonequilibrium steady states, and not just for small systems.  Manybody simulations of
time-reversible shear flows and heat flows likewise provide strange attractors with
fractional dimensionalities less than that of the phase spaces in which they are
embedded\cite{b10}. These attractors are ``strange'' [ fractal ] and ``chaotic''
[ because small perturbations on them {\it grow exponentially fast} ], despite the
continuous equations that generate them and despite their zero-volume attractive nature.
The dimensionality loss exhibited by attractors increases in an irregular manner
with the field-induced departure from equilibrium.

These fractal distributions are fully consistent with the Second Law of Thermodynamics.
That Law declares that only the dissipative forward-in-time versions of the
nonequilibrium trajectories are observable.  The time-reversed versions of the dynamics
-- unphysical trajectories which convert heat-reservoir energy into work -- are both
mechanically unstable and computationally unobservable.  Nonequilibrium dissipative
trajectories seek out fractal {\it attractors} when followed forward in time.  Reversed
trajectories make up an unstable unobservable phase-space {\it repellor}, a fractal
phase-space object which repels rather than attracts nearby trajectories and is unstable
to perturbations.  Picturing such a repellor in $[ \ \alpha,\sin(\beta) \ ]$ space is
easy.  Simply reflect the fractal attractor objects of Figure 2 about their horizontal
center lines. This changes the sign of the velocity at each collision and is equivalent to
picturing the motion backward in time.

The ubiquitous {\it fractal} nature of nonequilibrium steady states, singular everywhere,
indicates the difficulty inherent in attempting their mathematical description. Although
the motion of the hard-disk-scatterer Galton Board problem {\it is} ergodic for moderate
fields, with all states accessible, typically, for smoothly-continuous potentials  there
are {\it nonergodic} situations for simple mechanical systems. The mechanical treatment of
theoretical models with smooth potential minima is complicated by the complex phase-space
structure of Hamiltonian chaos -- ``islands'' ( the two-dimensional cross sections of
three-dimensional tori ),
chains of islands, . . . , and the endless details of structure on all scales, from large
to the microscopic and to the unobservably small\cite{b11}.

This unsettling complex situation can be avoided, or simplified, by the judicious use
of thermostating forces. The best problem area for such explorations is the one-dimensional
harmonic oscillator, the prototype of smooth Hamiltonian systems.  We will see how a
thermostated oscillator can be modeled so as to avoid the infinitely-many nonergodic
solutions of Hamiltonian mechanics while providing insight into irreversible processes
described by simple phenomenological laws. Let us turn next to the thermostating of
that simplest case, a single harmonic oscillator with a specified temperature $T$ rather
than a fixed Hamiltonian energy $E$ .

\section{Nos\'e Seeks Gibbs' Canonical Ensemble through Chaos}

Willard Gibbs invented his ``canonical'' ( in the sense of ``simplest'' or
``prototypical'' ) ensemble in order to link microscopic phase-space dynamics to
macroscopic temperature and thermodynamics.   His canonical ensemble collects
together all the energy states accessible to a system in contact with a heat
reservoir at a temperature $T$.  The relative weight of each such state in the
ensemble is the familiar Maxwell-Boltzmann weighting proportional to $e^{-E/kT}$ .
Here $k$ is Boltzmann's constant.

For simplicity we focus on the application of Gibbs' ensembles to the {\it one-dimensional}
harmonic oscillator.  With the mass and force constant and Boltzmann's constant all
set equal to unity Gibbs' canonical weighting of the oscillator states is the
familiar Gaussian distribution, a probability density for $q$ and $p$ :
$$
f(q,p) = (2\pi T)^{-1} e^{-q^2/2T}e^{-p^2/2T} \ .
$$
A common textbook rationalization of the canonical distribution is to imagine 
that the members of an ensemble of systems are weakly coupled to one another.
The coupling permits energy to be exchanged among the systems, resulting in
Gibbs' maximum-entropy canonical distribution. Shuichi Nos\'e developed a much
simpler picture in which a {\it single} system is coupled dynamically to a
computational heat reservoir in such a way that a long-time average of that system's
properties is identical to the canonical average.  Let us describe this idea in the
context of the one-dimensional harmonic oscillator.

\subsection{Nos\'e's Canonical Mechanics}
By 1984 Shuichi Nos\'e had documented his modification of Hamilton's constant-energy
dynamics in two ground-breaking papers\cite{b12,b13}.  His new dynamics was formally
consistent with Gibbs' constant-temperature canonical distribution.  For the
oscillator problem at the temperature $T$ , the simplest form of Nos\'e's novel
Hamiltonian, now with two degrees of freedom, $\# = 2$ , rather than one, has the
form :
$$
2{\cal H}_{\rm Nos\acute{e}}(q,p,s,\zeta) = (p/s)^2 + q^2 + \#T\ln(s^2) + \zeta^2 \ .
$$
The added thermostat degree of freedom, $s$ and its conjugate momentum
$\zeta \equiv p_s$ , along with the usual $(q,p)$ pair describes the canonical
oscillator problem with {\it four} ordinary differential equations rather than
the usual two :
$$
\dot q = (p/s^2) \ ; \ \dot p = - q \ ; \ \dot s = \zeta \ ; \ \dot \zeta = 
(p^2/s^3) - (\#T/s) \ . \ [ \ {\rm Nos\acute{e}} \ ]
$$

Nos\'e carried out the straightforward and tedious algebra necessary to show
that this approach can be made consistent with Gibbs' canonical distribution.
Three steps were involved in his demonstration : \\

\noindent
[ 1 ] ``Time Scaling'': $(d/dt)\equiv \ ^\cdot \longrightarrow s(d/dt)\equiv
 s\ ^\cdot $ ; \\
\noindent
[ 2 ] redefine momentum: $(p/s) \longrightarrow p$ ; \\
\noindent
[ 3 ] redefine degrees of freedom: $\# \longrightarrow \# - 1$ . \\

More than a decade passed before  Carl Dettmann simplified this approach\cite{b14,b15}.
He showed that starting out with a {\it scaled} Hamiltonian and setting it equal to zero ,
$$
{\cal H}_{\rm D} \equiv s{\cal H}_{\rm Nos\acute{e}} \equiv 0 \ ,
$$
produces a dynamics identical to Nos\'e's three-step approach without the need
for an explicit time-scaling step.

\begin{figure}
\includegraphics[width=5.25in,angle=90.]{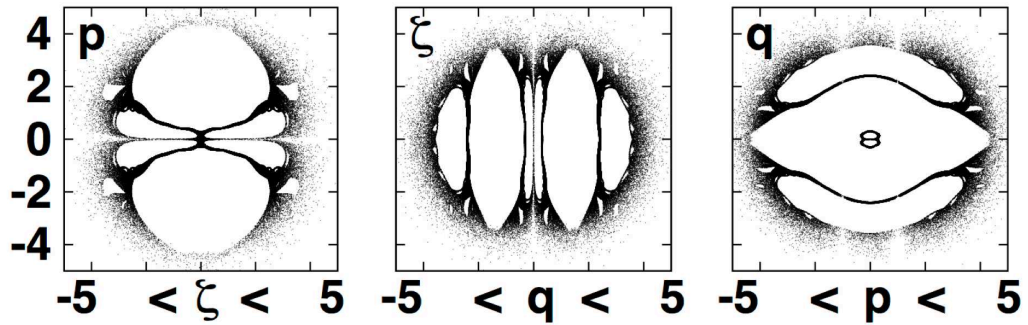}
\caption{
Trajectory intersections with the $q=0, \ p=0,$ and $ \zeta = 0$ planes are shown.
The points in these cross sections all correspond to penetrations by a single chaotic
trajectory with the initial conditions $(q,p,\zeta) = (0,5,0)$ .  The temperature is
unity for these three equilibrium Nos\'e-Hoover oscillator cross sections.
}
\end{figure}

\subsection{Nos\'e-Hoover Canonical Mechanics}
The even simpler ``Nos\'e-Hoover'' version of Nos\'e's approach\cite{b16} eliminates
all three of these steps as well as the extraneous variable $s$ . It is based on the
application of Liouville's phase-space continuity equation to the oscillator equations
of motion augmented by the definition of a time-dependent friction coefficient $\zeta$ :
$$
\dot q = p \ ; \ \dot p = -q -\zeta p \ ; \ \dot \zeta = [ \ (p^2/T) - 1 \ ] \ 
\longrightarrow \langle \ p^2 \ \rangle = T \ . \ [ \ {\rm NH} \ ]
$$
This friction coefficient acts as a ``thermostat'', steering the instantaneous
temperature $p^2$ toward the target thermostat temperature $T$ . We can verify that
this three-equation model is consistent with the canonical distribution for $q$ and
$p$ augmented by a Gaussian distribution for $\zeta$ :
$$
f(q,p,\zeta) \times (2\pi)^{3/2}T = e^{-q^2/2T}e^{-p^2/2T}e^{-\zeta^2/2} \ .
$$
To show this we evaluate the time-rate-of-change of the probability density
$f(q,p,\zeta)$  at a fixed location $( \ q,p,\zeta \ )$ in the three-dimensional
phase space where the local flow velocity is
$v = ( \ \dot q,\dot p,\dot \zeta \ )$ :
$$
(\partial f/\partial t) = -\nabla \cdot (fv) = -f(\partial \dot p/\partial p)
-(\partial f/\partial q)\dot q - (\partial f/\partial p)\dot p -
(\partial f/\partial \zeta)\dot \zeta =
$$
$$
f\zeta + (qf/T)(p) + (pf/T)(-q-\zeta p) + (\zeta f)[ \ (p^2/T) - 1 \ ] \equiv 0 \ .
$$
Because $(\partial f/\partial t)$ vanishes everywhere the Nos\'e-Hoover equations
{\it are} consistent with Gibbs' canonical distribution.

Although the smooth and simple three-dimensional Gaussian distribution is {\it the}
exact stationary solution of the Nos\'e-Hoover motion equations, the new dynamics
conceals an intricate complexity connected with ``chaos'', the exponential sensitivity
of calculated trajectories to perturbations of the initial conditions.  The three
[ NH ] motion equations included here are just the necessary minimum for chaos. But
these {\it necessary} three are not necessarily {\it sufficient}, as is easily
revealed by a numerical exploration of the oscillator's phase space distribution
$f(q,p,\zeta)$ .  Whether or not there {\it is} chaos depends, in a highly-complicated
way, upon the initial conditions. \\

To outline the chaos' profile, let us advance a trajectory beginning at a known
chaotic initial condition such as $( \ q,p,\zeta \ )$ = $( \ 0,5,0 \ )$ or
$( \ 3,3,0 \ )$ .  By plotting any two of the three variables $\{ \ q,p,\zeta \ \}$
just as the third passes through zero, three separate cross sections of the chaotic
sea are revealed. {\bf Figure 4} shows the sequence of 768,460 such successive
crossing points following from the initial condition
$\{ \ q,p,\zeta \ \} = \{ \ 0.00,5.00,0.00 \ \}$
using one billion fourth-order Runge-Kutta timesteps with $dt = 0.001$ .  We see from
{\bf Figure 4} that only about six percent of the three-dimensional stationary Gaussian
measure makes up the chaotic sea.  Any trajectory started in the sea cannot leave and
eventually explores all of it.  

\begin{figure}
\includegraphics[width=5in,angle=+90.]{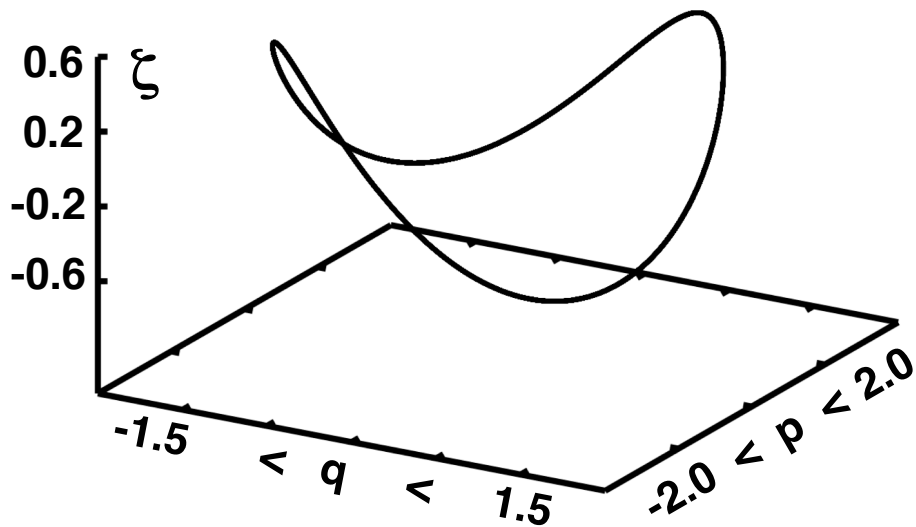}
\caption{
Simplest [ shortest period ] periodic orbit for the Nos\'e-Hoover oscillator with $T = 1$ .
The initial condition is the turning point
$(q,p,\zeta) = (0.0,1.5499,0.0)$ . The motion equations are
$\dot q = +p \ ; \ \dot p = -q -\zeta p \ ; \ \dot \zeta = p^2 - 1$ .
The $(q,p/s,\zeta)$ trajectory from the original Nos\'e equations with $s$ initially
equal to unity and with $\#$ reduced from 2 to 1 [ so that the motion equations are
$\dot q = (p/s^2) \ ; \ \dot p = -q \ ; \ \dot \zeta = (p^2/s^3) - (1/s)$ ] traces out
exactly the same $(q,p,\zeta)$ trajectory pictured here, but with a period 7.1973 rather
than 5.5781 .  The Nos\'e-Hoover Lyapunov exponent varies in the range $\pm0.6513$ on this
orbit.
}
\end{figure}

The remaining phase space is occupied privately by concentric tori enclosing stable
periodic orbits. The simplest such orbit for the Nos\'e-Hoover oscillator is shown
in {\bf Figure 5}. The repeat time for this orbit is 5.5781.  It includes four
symmetric turning points :
$$
\{ \ q,p,\zeta \ \} = \{ \ 0.0000,\pm 1.5499,0.0000 \ \} \ {\rm and} \
\{ \ \pm 1.2144,0.0000,0.0000 \ \} \ .
$$
This orbit is the central core of an infinite family of nested tori.  See {\bf Figure 6} .
The tori are quasiperiodic regular structures tracing out two-dimensional private
regions where there is none of the chaos present which would make new three-dimensional
regions accessible. Let us have a more detailed look at the mechanism enabling chaos by
measuring the {\it rates} at which chaotic orbits separate on their exploratory journeys.

\begin{figure}
\includegraphics[width=5in,angle=-90.]{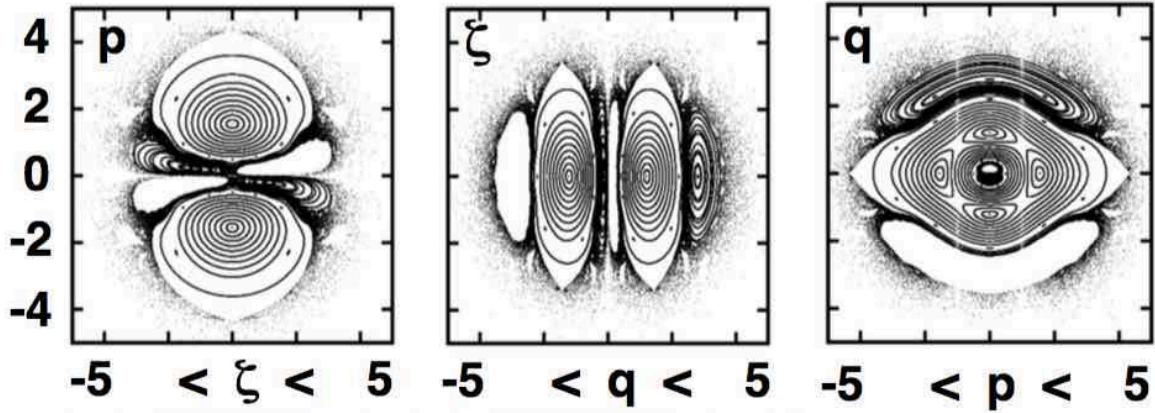}
\caption{
Cross sections with $q=0,$ $p=0$, $\zeta=0$, starting with the periodic orbit penetration
at $(\pm 1.2144,0,0)$ and increasing or decreasing the initial coordinate value of
$\pm1.2144$ by 15 successive steps of $\pm0.1$, ending up with the initial $q$ values
$\pm 2.7144$ .   The resulting cross-sections are shown. The 12th positive and negative
initial conditions lie within the chaotic sea.  All the rest generate
tori which trace out individual points along the lines shown in this figure. 
}
\end{figure}

\subsection{Characterizing Chaos through Lyapunov Instability}

Chaos is an important topic.  Without it there would be no hope for correspondence
between the microscopic and macroscopic descriptions of material behavior. Chaotic
trajectories exhibit ``Lyapunov instability'' --- two nearby trajectories [ a
``reference'' and its ``satellite'', as explained below ] rapidly
separate.  When averaged over the chaotic sea ( that both of them inhabit ) the mean
value of this instantaneous separation rate defines the largest ( positive )
Lyapunov exponent, $\lambda_1$ .  This instantaneous separation rate ,
$\lambda_1(t) = | \ \dot r_1 - \dot r_2 \ |/| \ r_1-r_2 \ |$ ,
is an observable which is easy to measure.

Giancarlo Benettin's Lyapunov-exponent algorithm\cite{b17} follows the simultaneous
dynamics of a ``reference'' trajectory and a nearby ``satellite'', rescaling their
separation to maintain their closeness at the end of every timestep. Typically only
the position of the satellite is adjusted to restore the length of the separation
vector $| \ r_s -r_r \ |$ .  Alternatively both of two neighboring trajectories
$r_1$ and $r_2$ can be adjusted symmetrically. Evidently a theoretical treatment
becomes complicated at the boundary where toroidal and chaotic trajectories meet,
mix, and coexist.   A more elegant continuous version of Benettin's step-by-step
rescaling can be implemented by including a constraining Lagrange multiplier in the
differential equations of motion\cite{b18}.  In the symmetric case the multiplier is applied
to both trajectories :
$$
\dot r_1 = v_1 + (\lambda/2)(r_2 - r_1) \ ; \
\dot r_2 = v_2 + (\lambda/2)(r_1  -r_2) \ .
$$
Here $v$ represents the unconstrained motion equations while
$\dot r = ( \ \dot q,\dot p,\dot \zeta \ )$ describes the actual constrained motion.
The Lagrange multiplier enforcing the constraint of fixed separation is just the
local value of the largest Lyapunov exponent.  Its longtime average {\it is} the
largest of the three global Lyapunov exponents which together constitute the
``Lyapunov spectrum'' of the three-equation Nos\'e-Hoover oscillator :
$$
\lambda_1 = \langle \ \lambda_1(t) \ \rangle  \ ; \
\lambda_1(t) = (v_1-v_2) \cdot (r_1-r_2)/(r_1-r_2)^2 \ .
$$
$\lambda_2$ and $\lambda_3$ are defined in terms of the comoving
growth or decay rates of an {\it area} defined by three trajectories --- the rate
is $\lambda_1 + \lambda_2$ ; and of an infinitesimal {\it volume} $\otimes$
defined by four trajectories with rate :
$$
(\partial \dot q/\partial q) + (\partial \dot p/\partial p) +
(\partial \dot \zeta /\partial \zeta) = 0 - \zeta(t) + 0
\equiv (\dot \otimes/\otimes) =
\lambda_1(t) + \lambda_2(t) + \lambda_3(t) \ .
$$ 

For the isothermal Nos\'e-Hoover oscillator with all the parameters equal to unity
that equilibrium oscillator, chaotic or not, has no dissipation.  This is implied
by its close relationship to the {\it four} oscillator equations of motion according
to Nos\'e's Hamiltonian formulation :
$$
\dot q = (p/s^2) \ ; \ \dot p = - q \ ; \ \dot s = \zeta \ ; \ \dot \zeta = 
(p^2/s^3) - (\#T/s) \ . \ [ \ {\rm Nos\acute{e}} \ ]
$$

Whether \# is chosen equal to 1 or to 2, the four Lyapunov exponents describing Nos\'e's
oscillator sum to zero as a consequence of Liouville's Theorem :
$$
(\partial \dot q/\partial q) + (\partial \dot p/\partial p) +
(\partial \dot s/\partial s) + (\partial \dot \zeta /\partial \zeta)
\equiv (\dot \otimes/\otimes) =                   
\lambda_1(t) + \lambda_2(t) + \lambda_3(t) + \lambda_4(t) \equiv  0 \ .
$$

The mean value of the friction coefficient,  $\langle \ \zeta \ \rangle$ , vanishes in
both cases [ Nos\'e or Nos\'e-Hoover ] because the motion equations are ``conservative''
whether or not the initial condition is chaotic.  Accordingly, although the local
Nos\'e-Hoover spectrum sums to $-\zeta(t)$ its time-averaged spectrum sums to zero :
$$
\langle \ (d\ln \otimes /dt) \ \rangle = \langle \ (\partial \dot p/\partial p) \
\rangle = \langle \ -\zeta \ \rangle = \lambda_1 + \lambda_2 + \lambda_3 = 0 \ .
$$

In view of Hamiltonian mechanics' time-reversibility both spectra are also
{\it symmetric}, with the first and last time-averaged exponents summing to zero.
Within the chaotic sea the Lagrange-multiplier analyses of the equilibrium oscillators
give :
$$
\{ \ \lambda \ \} = \{ \ \lambda_1,\lambda_2,\lambda_3,\lambda_4 \ \} =
\{ \ +0.001925,0.000,0.000,-0.001925 \ \} \ ; \ [ \ {\rm Nos\acute{e}} \ ] 
$$
$$
\{ \ \lambda \ \} = \{ \ \lambda_1,\lambda_2,\lambda_3 \ \} =
\{ \ +0.0139,0.000,-0.0139 \ \} \ . \ [ \ {\rm Nos\acute{e}-Hoover} \ ]
$$
In the quasiperiodic toroidal regions where chaos is absent  {\it all} the
time-averaged Lyapunov exponents are zero.  Although the time averaging produces
symmetric spectra the instantaneous spectra need not be symmetric.  Let us
demonstrate that somewhat surprising lack of forward-backward symmetry next.

\subsection{The Tricky Time Reversibility of the Nos\'e-Hoover Lyapunov Spectrum}

One might well expect that the {\it ``local''}  ( instantaneous ) oscillator
Lyapunov spectra are time-reversible too.  After all, both the trajectories used
to define the largest Lyapunov exponent {\it are} reversible.  Growing separation,
forward in time, corresponds to diminishing separation when reversed, and
{\it vice versa}.  A simple way to test this idea of time-reversible Lyapunov
exponents is to store a reference trajectory $\{ \ q,p,\zeta \ \}$ going forward
in time, with $dt > 0$ .  Then, analyzing the instabilities forward and backward in
time, as described by the tendency of an adjustable ``satellite'' trajectory to flee
or approach a stored ``reference'' is a fruitful approach.

Time reversibility can be imposed on a stored reference trajectory $\{ \ q,p,\zeta \ \}$
in either of two ways.  The stored data can be used ``as is'' for a reversed
trajectory simply by changing the sign of the time increment $dt$ .  Then the stored
data are obviously solutions of the three reversed motion equations :
$$
\dot q = -p \ ; \ \dot p = +q +\zeta p \ ; \ \dot \zeta = 1 - p^2 \ {\rm for} \
dt < 0 \ .
$$
Alternatively one can simply change the signs of the stored $\{ \ p \ \}$ and
$\{ \ \zeta \ \}$ [ and $\{ \ \lambda \ \}$ if the Lagrange multiplier approach
is used ] .  Then the modified sequence of $\{ \ q,p,\zeta \ \}$  satisfies the
original motion equations when it is played backward.  These two ways of exploring
reversibility are precisely equivalent in computation.  Only the signs of some
of the numbers are changed.  The magnitudes forward and backward are identical.

If one were to follow this reversibility procedure for {\it  both} the reference
and the satellite trajectories, then increasing separation going forward in time
would correspond to decreasing separation in the reversed trajectory, and hence
to a negative Lyapunov exponent rather than a positive one.  Evidently in the
Nos\'e-Hoover oscillator case, with just three exponents, the largest exponent
going forward would become the most negative going backward :
$$
\lambda_1^f = \ \langle \ \lambda_1(t) \ \rangle_{dt>0} =
- \ \langle \lambda_3(t) \ \rangle_{dt<0} \ .
$$
Although the Lagrange-multiplier equations using a stored reference trajectory
with stored satellite trajectories {\it are} all of them time-reversible, with
both $v(t)$ and $\lambda_1(t)$ changing sign if $dt<0$ , the reversed exponents
are erroneous !  The correct approach, storing a reference trajectory and then
generating a {\it new} set of reversed-time satellite trajectories to go with the
reference, using Lagrange multipliers or with Benettin's
rescaling algorithm, gives an instantaneous spectrum which is unrelated to its
forward-in-time twin.  For us this seems surprising, even though it is fairly
well known, and suggests interesting research directions, perhaps leading to
a better understanding of which spectra reverse and which do not.  The simplest
explanation is probably best : the tendency of two trajectories
to separate can depend only on their {\it past} history, and not the future.  On a
particular time-reversible trajectory, $\{ \ q(ndt) \ \}$ , without knowing
whether $dt$ is positive or negative, the notions of past and future could be
thought nebulous.  The low-cost readily-available cure for this uncertainty is
straightforward computation and observation, finding out what {\it does} happen.

\begin{figure}
\includegraphics[width=3in,angle=-90.]{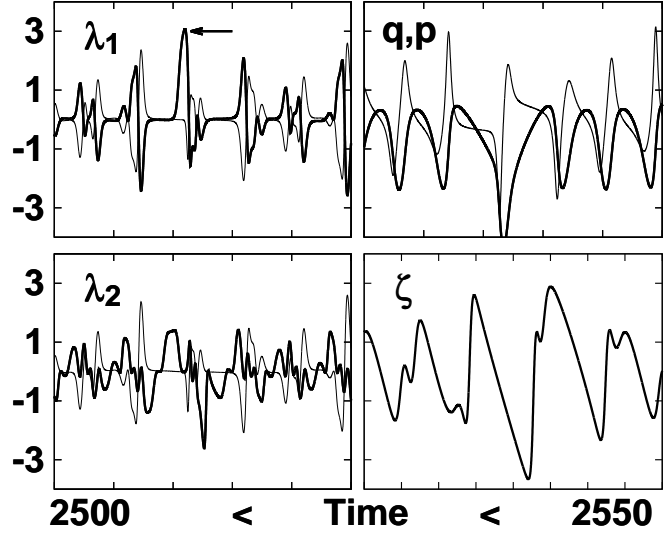}
\caption{
The summed spectra, both forward and backward, and the negative
value of $\zeta$ all correspond to the same bold curve in the lower right panel.
The left panels represent the first two  Lyapunov exponents forward ( thick lines )
and backward ( thin lines ) in time.  The summed exponents, forward and
backward for the two methods of time reversal, satisfy similar sum rules :
$                                                                                                 
\lambda_1 + \lambda_2 + \lambda_3 = \mp \zeta \ .                                                   
$
The coordinate $q$ (thick line) and the momentum $p$ (light line) versus time are
shown in the upper right panel.  Notice that a brief segment of time, just past
2520, during which the momentum is near zero corresponds to the only part of the
plotted interval in which the first Lyapunov exponents have very different
magnitudes in the forward and backward directions.  The arrow points out the maximum
in $\lambda_1^f$ to which $\lambda_1^b$ is unrelated.
}
\end{figure}

What actually {\it does} happen is surprising, and is illustrated in
{\bf Figure 7} . The largest Lyapunov exponent going forward in time and the
largest going backward typically sum nearly to zero, indicating that the vectors
joining the reference and satellite trajectories are almost parallel in the two
time directions. The stretching or shrinking observed for $dt > 0$ is replaced
by its opposite, shrinking or stretching for $dt < 0 $ :
$$
+\lambda_1^{dt>0}(t) \simeq -\lambda_1^{dt<0}(t) \ .
$$ 
Of course this cannot be precisely true as the averaged values of both versions
of $\lambda_1$ are positive.  But the fluctuations in the exponent are two
orders of magnitude larger than the relatively-small time-averaged value of
$\pm 0.0139$ .  The fluctuations are of the order of the oscillator frequency,
$\omega = 2\pi \nu$ , rather than the much smaller instability rate.

A trajectory fragment [ 50,000 timesteps forward and backward from near the center
a much longer  simulation, $0 < t < 5000$ ] and specially selected to show the
reversibility  phenomena, is analyzed in {\bf Figure
7} . Apart from the single strong maximum in $\lambda_1(t)^{dt>0}$ indicated by
the arrow the sum of the two exponents is close to zero. There is no apparent
correlation between the values of the second or the third Lyapunov exponents
in the two time directions. Thus trajectory stability depends qualitatively on
the direction of time, even for equilibrium systems.

In any case, the sums of all three exponents going forward and backward in time
both must obey the instantaneous sum rule :
$$
\lambda_1(t) +\lambda_2(t) +\lambda_3(t) = (d\ln\otimes/dt) = \mp \zeta\ .
$$
The equilibrium oscillator considered here behaves like a dissipative
system with an asymmetry between the forward and backward directions of time
which can be traced to the varying comoving phase volume $\otimes(t)$.  With even
a little thermostating the resulting heat transfer is enough to break the 
pairing symmetry expected for Hamiltonian systems.  In nonequilibrium steady
states we will see that this same symmetry breaking is unrelenting and in fact
prevents reversing the formally time-reversible trajectories by any means other
than reusing stored trajectories.

\section{Oscillator Ergodicity {\it via} Generalized Friction}

We have seen that numerical solutions of the three Nos\'e-Hoover oscillator equations
$( \ \dot q,\dot p,\dot \zeta \ )$ are far from ergodic.  The same is true for the
four Nos\'e equations $( \ \dot q,\dot p,\dot s,\dot \zeta \ )$ because the two
trajectories are {\it identical} for the corresponding scaled variables :
$$
\dot t_{NH} \equiv s\dot t_{N} \longrightarrow
\{ \ q,p,\zeta,t \ \}_{NH}
\equiv \{ \ q,(p/s),\zeta,t \ \}_{N( \ {\rm scaled} \ )} \ .
$$
Provided that the initial values of $\{ \ q,p,\zeta \ \}$ correspond at
$t_{NH} \equiv t_N \equiv 0$ with $s(t=0)\equiv 1$ plots of $\zeta(q)$ are identical
for the two sets of differential equations.  This equivalence is a useful demonstration
of Nos\'e's ``scaling of time'' with the time-scaling variable $s$ .

Though ergodicity is lacking, the friction coefficient equations for $\dot \zeta$
guarantee that the second velocity moments are equal to the specified value of
the temperature$$
\langle \ p^2 \ \rangle_{NH} \equiv \langle \ (p/s)^2 \ \rangle_{N} \equiv T \ .
$$
By controlling another moment the velocity distribution should come more closely to
resemble the equilibrium Maxwell-Boltzmann Gaussian.  In fact two moments can be
enough\cite{b19}. Consider the simultaneous control of $p^2$ and $p^4$ . For this
generalized Nos\'e-Hoover oscillator problem {\it two} friction coefficients are
involved, $\zeta$ for $p^2$ and $\xi$ for $p^4$ :
$$
\dot q = +p \ ; \ \dot p = -q -\zeta p -\xi (p^3/T)\ ; \
\dot \zeta \propto (p^2/T) - 1 \ ; \ \dot \xi \propto (p^4/T^2) - 3(p^2/T)
\ . \ [ \ {\rm HH} \ ]
$$
Solutions of these ``Hoover-Holian'' [ HH ] equations, with all parameters equal
to unity, appear to be ergodic, with the probability density and {\it all} of its
moments converging to Gibbs' Gaussian moments better and better as the sampling
time is increased. At the same time careful examination of the two-moment flow's
cross sections reveals that there are no ``holes'' in the {\it doubly-thermostated}
chaotic sea. This is in marked contrast to the holy-sea sections seen in the
{\it singly-thermostated} flows illustrated in  {\bf Figures 4 and 6} .
Comprehensive tests for ergodicity were formulated and applied to the
doubly-thermostated [ HH ] equations in connection with the 2014 Snook
Prize\cite{b20}.

Evidently one can never achieve {\it perfect} agreement in numerical tests as
the probabilities of high-energy states are not only infinitesimal but would
also require infinitesimal timesteps for computational stability. The four
[ HH ] equations above, which control the second and
fourth velocity moments, are only one of several methods for achieving ergodicity
for the oscillator with the use of {\it two} friction coefficients.  In view of this
success it is natural to wonder whether or not a {\it single} carefully-chosen
thermostating variable could make the dynamics ergodic.

There are many different approaches to the ergodicity problem.  The symmetry of the
oscillator's coordinate and momentum suggests that one could thermostat $q^2$ just
as well as $p^2$ or perhaps even both.  By interpreting $q^2$ as the fluctuation of
the force these ideas can be, and have been, applied to more complicated systems.  
Extensions of this idea to the ``weak'' ( time-averaged ) control of two or more
different moments, with forces proportional to
$$
[ \ (q^2/T) - 1 \ ], \ [ \ (q^4/T^2) - 3(q^2/T) \ ], \ [ \ (p^2/T) - 1 \ ],
 \ [ \ (p^4/T^2) - 3(p^2/T) \ ], \ .\ . \ . \ ,
$$
have proved fruitful.  Rather than describe all of these efforts let us concentrate
on the simplest of them, the simultaneous weak control of both $(p^2/T)$ and $(p^4/T^2)$
{\it using a single friction coefficient}. A useful tool in investigations of this sort
is the $\chi^2$ test, which makes it possible to estimate numerically  which of two
different thermostat choices gives results closer to the Maxwell-Boltzmann velocity
distribution. We turn to that next.

\subsection{Ergodicity with a Single Thermostat Variable -- the $\chi^2$ Test}

After 30 years of failed attempts involving dozens of investigators we recently
discovered that ergodic control of a harmonic oscillator {\it appears to be} possible
with just a single thermostat variable.  An important clue, from the investigations of
Bulgac, Kusnezov, Ju, and Bauer\cite{b21,b22}, was that {\it cubic} thermostats provide
enhanced ergodicity relative to linear ones.  With cubic thermostats ergodic motion can
occur even in the minimal case of a three-dimensional phase space.  For maximum
simplicity we assign any and all control variables to the momentum because the canonical
momentum distribution is always the same, independent of the chosen potential energy.

Accordingly, let us assign the entire thermostating burden to the momentum through
just two velocity moments, again using the harmonic oscillator to illustrate :
$$
\dot q = p \ ; \
\dot p = -q - \zeta^n[ \ \alpha p + \beta (p^3/T) + \gamma (p^5/T^2) \ ] \ ;
$$
$$
\dot \zeta = \alpha[ \ (p^2/T) - 1 \ ] + \beta[ \ (p^4/T^2) - 3(p^2/T) \ ] 
+ \gamma[ \ (p^6/T^3) - 5(p^4/T^2) \ ] \ .
$$
With this functional form of control, where $n=1$ or 3 , it is easy to show that the
corresponding solution of Liouville's continuity equation is again the Maxwell-Boltzmann
Gaussian distribution :
$$     
f(q,p,\zeta) = e^{-q^2/2T}e^{-p^2/2T}e^{-\zeta^{n+1}/(n+1)} \ .
$$
Monte Carlo or grid-based exploration of $(\alpha,\beta,\gamma)$ parameter space
reveals many binary combinations of two types, $(\alpha,\beta,0)$ and
$(\alpha,0,\gamma)$ , which fill out Gibbs' distribution. Candidate models
have good velocity and coordinate moments as well as phase-space cross sections 
without discernable ``holes''.  {\bf Figure 8} shows $\zeta = 0$
cross sections for two successful combinations $( \ \alpha,\beta,\gamma \ ) =
( \ 0.05,0.32,0.00 \ )$ and $( \ 1.50,0.00,-0.50 \ )$.  Because motion equations
involving the fifth and sixth powers of velocity are ``stiff'' we have mostly
restricted our detailed explorations to combinations of the second-moment and
fourth-moment controls.

How best to find such combinations ?  There are many ways.  A purely visual
approach is relatively effective.  A hundred-frame movie is a convenient
visualization tool. The frames result from choosing an arbitrary grid of 100
$( \ \alpha,\beta \ )$ values and plotting 200,000 successive $( \ q,p \ )$
cross-section points for which $\zeta$ vanishes.  Frames lacking apparent holes
in the density, and also providing good second, fourth, and sixth moments identify
$( \ \alpha,\beta \ )$ choices as good candidates for ergodicity.  The most
interesting sections can be confirmed ergodic with greater accuracy by zooming in,
while  using a few million section points rather than 200,000 .

\begin{figure}
\includegraphics[width=5in,angle=+90.]{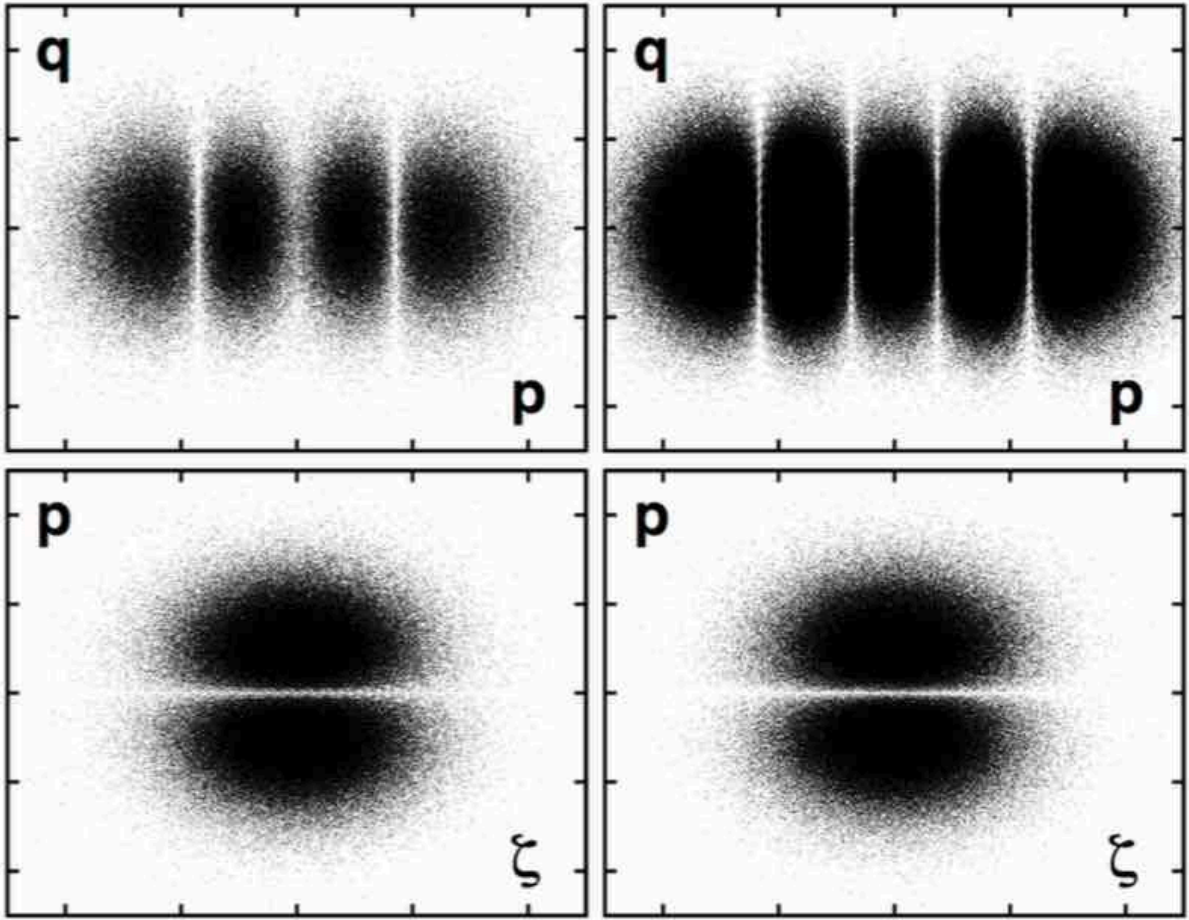}
\caption{
Cross sections for $( \ \alpha,\beta,\gamma \ ) =
( \ +0.05,+0.32,0.00 \ )$ [ at the left ] and $( \ 1.50,0.00,-0.50 \ )$ 
[ at the right ].  These sections are for equilibrium oscillators $[ \ T = 1 \ ]$
and use controls linear and cubic in $p$ while linear in $\zeta$ , with $n=1$ .  The white lines
correspond to ``nullclines'' where the velocity normal to the cross section vanishes. 
The scales all range from $ -5.0 \ {\rm to} \ +5.0$ . The upper panels are the 
$\zeta = 0.0$ sections.  The lower panels are the $q = 0.0$ sections.
}
\end{figure}

A more systematic approach, also useful, but not at all foolproof, can be based on
Pearson's $\chi^2$ statistic. $\chi^2$ comes in handy when it is desired to know
if an {\it observed} distribution $\{ \ o \ \}$ ( from a numerical simulation )
is consistent with a predicted one ( with {\it expected} values $\{ \ e \ \}$ from a
theoretical analysis of the flow equations ).  For a candidate $( \ \alpha,\beta \ )$ 
combination a coarse-grained probability can be defined and determined within $N$ discrete
sampling bins. The mean-squared deviation of the bin probabilities from the expected
theoretical value is divided by the expected bin population and averaged over bins.
The Central Limit Theorem predicts that the resulting sum :
$$
\chi^2 \equiv \sum^N \langle \ (o - e)^2/e \ \rangle \ ,
$$
where the angle-bracket average is over the points in each of the bins, will
approach the number of bins $N$ if the ``expected'' already-known distribution
$\{ \ e \ \}$ , is a match to the ``observed'' one, $\{ \ o \ \}$ .

The simplest special case of this idea results if a uniformly flat distribution of random
numbers $\{ \ 0 < {\cal R} < 1 \ \}$ is divided into $N$ equal sampling bins.  With just
two bins the large-sample limit is $\chi^2 =$  1 ; with four bins 3 ; with six bins 5,
and so on. The limiting value of $\chi^2$ with $N$ bins is just $N-1$ . With data gathered
from dynamical simulations where the distribution is {\it not} flat the dependence on the sample
size is irregular and the convergence is slowed due to the inevitable serial correlation
of sampled trajectory data.

Knowing that Gibbs' canonical oscillator distribution is Gaussian in all the variables
makes it possible to test bins in $q$ or $p$ or $\zeta$, or combinations of these
variables, using the $\chi^2$ goodness-of-fit criterion. With 100 bins and a billion
data points values of $\chi^2$ within ten percent of the number of bins, $\chi^2 \simeq
100 \pm 10$ , are typical when the distribution being observed really {\it is} Gaussian.
It is possible to debug such a program using a good random number generator such as
Press' ``ran2'' generator from {\it Numerical Recipes}.

Our investigations suggested that control variables based on the differential
equations :
$$
\dot q = p \ ; \ \dot p = -q - \zeta^n[ \ \alpha p + \beta (p^3/T) + \gamma(p^5/T^2) \ ] \ ; \
$$
$$
\dot \zeta = \alpha[ \ (p^2/T) - 1 \ ] + \beta[ \ (p^4/T^2) - 3(p^2/T) \ ] +
\gamma[ \ (p^6/T^3) - 5(p^4/T^2) \ ] \ ,
$$
{\it can} provide ergodicity for the harmonic oscillator.  The equations are particularly
stiff when the sixth moment is included so that $(\alpha,\beta) = (0.05,0.32)$ is the most
promising of the combinations we have tried to date. We have also included sample sections
 using a two-parameter $( \alpha,\gamma)$ model controlling the second and sixth velocity
 moments.  Because these equations {\it are} stiffer, requiring timesteps of order 0.0001
rather than 0.001, the second and fourth-moment control is the better choice. With this model
the fluctuations of the largest local Lyapunov exponent correspond to a standard deviation
of about 27, two orders of magnitude larger than its long-time-averaged value so that the
exploration of phase space is relatively rapid. 

Because the equilibrium Maxwell-Boltzmann momentum distribution $e^{-p^2/2T}$ is the
{\it same} for any potential, this finding suggests that it may well be possible to
thermostat {\it any} small Hamiltonian system in this way.  Let us check this idea for the
simple pendulum.

\subsection{Thermostating the  Nos\'e-Hoover Pendulum Problem}

We have seen that the Nos\'e-Hoover oscillator can be thermostated in a wide variety
of ways using either one, two ( or possibly even three ! ) control variables, though
the stiffness suggests that using  three controls is unwise.  In each case the consistency
of the solutions can be checked using
Liouville's Theorem to confirm that the stationary flow leaves Gibbs' probability
density unchanged :
$$
(\partial f/\partial t) = -\nabla \cdot (fv) = 0 \ .
$$                                        

To apply similar ideas to the pendulum problem\cite{b23} using a single friction coefficient
we need only to replace the potential energy: $(q^2/2) \rightarrow -\cos (q)$ so that Gibbs'
canonical distribution becomes :
$$
{\cal H} = -\cos(q) + (p^2/2) \rightarrow f(q,p,\zeta) \propto
e^{\cos(q)/T}e^{-p^2/2T}e^{-\zeta^{n+1}/(n+1)} \ {\rm with} \ -\pi < q < +\pi \ . 
$$
We can make the friction coefficient $\zeta$ consistent with the Gaussian
momentum distribution by using an arbitrary collection of moments, for instance :
$$
\dot q = p \ ; \
\dot p = -\sin(q) -\zeta^n[ \ \alpha p + \beta (p^3/T) + \gamma (p^5/T^2) \ ] \ ; \
$$
$$
\dot \zeta = \alpha[ \ (p^2/T) - 1 \ ] + \beta[ \ (p^4/T^2) - 3(p^2/T) \ ] +
\gamma[ \ (p^6/T^3) -5(p^4/T^2) \ ]  \ .
$$
The cross sections for $(\alpha,\beta) = (0.300,0.300)$ [ found visually ]
and $(0.088,0.188)$ [ found using Pearson's $\chi^2$ test ] are shown in the {\bf Figure 9}
and, at least from the visual standpoint, both distributions appear to be ergodic. The
color indicates the magnitude of the local Lyapunov exponent, $\lambda_1(t)$ .

Our oscillator and pendulum examples both suggest that there {\it is} a dynamics,
deterministic and time-reversible, which closely follows Gibbs' canonical distribution.
Because many of the problems addressed with molecular dynamics involve isothermal
rather than isoenergetic processes this makes isothermal molecular dynamics a
particularly useful tool.

With weak polynomial control of the momentum it appears that we have completed Nos\'e's
search for a deterministic time-reversible dynamics satisfying Gibbs' canonical
distribution.  With equilibrium thermostats under control let us go on to consider
the extension of this dynamics to {\it nonequilibrium} systems in which the temperature
varies with location\cite{b8}.  We demonstrate the possibilities by including a temperature
gradient in the harmonic oscillator problem, allowing the dynamics to become dissipative
by transferring heat from ``hot'' to ``cold''. These problems have an intrinsic
pedagogical interest because they are simultaneously time-reversible {\it and}
dissipative. They generate multifractal attractor-repellor pairs, often with a
considerable \ae sthetic interest.

\begin{figure}
\includegraphics[width=5in,angle=+90.]{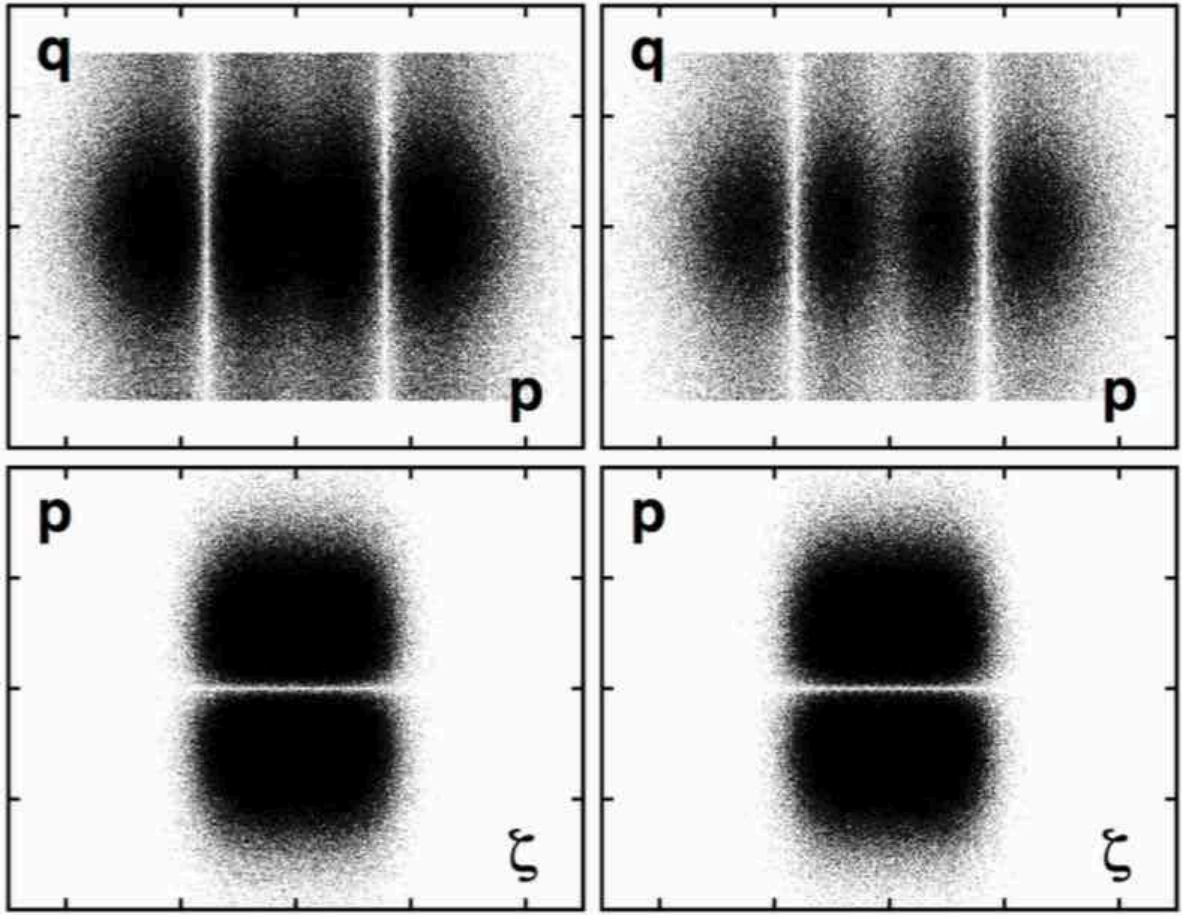}
\caption{
Ergodicity of the isothermal Simple Pendulum with $T=1$ and cubic control
[ $n=3$ and $\dot p \propto -\zeta^3$ ] .
These cross sections correspond to $(\alpha,\beta)=(0.300,0.300)$ 
to the left and $(0.088,0.188)$ to the right.  $\gamma = 0$ in both cases.The vertical scales
range from $-4.0$ to $+4.0$ and the horizontal scales range from
$-5.0$ to $+5.0$ .  Note that the range of the pendulum
coordinate ( where $q$ is an angle ) is periodic : $-\pi < q < +\pi$ .
}
\end{figure}

\section{{\it Nonequilibrium} Time-Reversible Dissipative Oscillators}

Gibbs' equilibrium canonical distribution depends upon the temperature
$T = \langle \ p^2/mk \ \rangle$
where $p$ is a cartesian momentum component for a particle with mass $m$ .  For simplicity
we continue to choose both $k$ and $m$ equal to unity.  We can introduce a {\it nonequilibrium}
temperature gradient, $\nabla T$, by choosing a coordinate-dependent temperature $T(q)$ . This
opens up the possibility for heat transfer leading to a quantitative treatment of nonequiibrium
problems.  We choose a smooth profile with a maximum temperature gradient $\epsilon$ :
$$
T(-\infty) =
1 - \epsilon <  T(q) \equiv 1 + \epsilon \tanh(q) < 1 + \epsilon 
= T(+\infty) \ . 
$$
Although the equations of motion ,
$$
 \dot q = p \ ; \ \dot p = -q - \zeta p \ ; \ \dot \zeta = \dot \zeta(p^2,p^4,p^6,T) \ ,
$$
are still time-reversible [ with $(+t,+q,+p,+\zeta) \rightarrow (-t,+q,-p,-\zeta)$ ]
the dynamics {\it can} turn out to be {\it dissipative} and {\it irreversible} .  How can
this be? What does it mean?  To address these questions, which are good ones, we must look
at time reversibility in more detail.

The concept of time reversibility\cite{b24} can be made unnecessarily complex by introducing
the concept of phase-space involutions.  A straightforward definition is the wiser choice:
First, imagine a {\it movie} of the motion in question ( this presupposes a connection
between the dynamical system of differential equations and objects capable of visual
representation ) ; Second, play the movie backwards, ( but with the clock on the wall
still recording a steady increase in the time ) just reversing the order of the
frames.  In the backward movie velocities change sign but coordinates do not.  If the
backward movie obeys {\it the same} equations as the forward one the dynamical system
describing the motion is time-reversible.  If not, then not.  Variables odd in the time,
such as velocity and the microscopic heat flux, change sign in the reversed motion, but
parameters, like gravity, are held fixed.  In every case that we study here our
microscopic differential equations of motion satisfy this criterion. In the simplest
example, the Nos\'e-Hoover oscillator ,
$$
\dot q = +p \ ; \ \dot p = -q -\zeta p \ ; \ \dot \zeta \propto [ \ (p^2/T) - 1 \ ] \ ;   
$$
reversing the sign of $p$ and the variable $\zeta$ is equivalent to reversing the
sign of the time so that this system {\it is} time-reversible, even in the case
where the temperature $T$ depends on the coordinate $q$ .  In the simplest
nonexample,
$$
\dot q = +p \ ; \ \dot p = -q -\zeta p \ ,
$$
where $\zeta$ is now a fixed parameter, not a variable, changing the sign of p
corresponds to reversing the $\dot q$ equation but {\it not} the $\dot p$ one
so that the constant-friction system is {\it not} time-reversible.  The same is true
of continuum solutions of viscous fluid flows and Fourier heat flow.  The continuum
constitutive relations for the shear stress of a Newtonian fluid with a velocity
gradient and the heat current in a Fourier heat conductor with a temperature
gradient :
$$
\sigma_{xy} = \eta[ \ (\partial u_x/\partial y) + (\partial u_y/\partial x) \ ] \ ; \
Q_x = -\kappa (\partial T/\partial x) \ ,
$$
are specially interesting. Here $\eta$ is the shear viscosity and $\kappa$ is the
thermal conductivity.  From the virial theorem we know that shear stress
$\sigma$ is an even ( time-reversible ) function of the velocities, while the
heat flux vector $Q$ is odd.  Both these observations contradict the
phenomenological macroscopic constitutive relations laid down by Newton and Fourier.

From thermodynamics we are well aware that ``The entropy of the Universe increases.''
Entropy is associated with heat reservoirs.  When a reservoir absorbs heat
$\delta Q$  its entropy increases by $\delta Q/T$ .  Likewise, when it releases heat
the reservoir entropy decreases.  Let us get back to the main question, ``How do
time-reversible motion equations produce irreversible behavior ?''

It is a curious and hard-to-grasp fact that reversible mathematical equations can
lead to irreversible behavior in the presence of Lyapunov instability, when the
separation between two nearby trajectories increases.  There would seem to be no
reason why an increasing separation would win out over a decreasing one,
particularly in the case where phase volume is conserved.  But a positive Lyapunov
exponent signals the system's seeking out the direction of increased phase-space
states.  Think again of computing the largest Lyapunov exponent by rescaling the
separation of two nearby trajectories.  That is, consider the linear variation of
a coordinate perturbation backward and forward in time, $\delta$ , paralleling the
direction characterized by the local Lyapunov exponent $\lambda(t)$ :
$$
\dot \delta/\delta = \lambda(t) \rightarrow
\langle \ \delta(t)/\delta(0) \ \rangle = (1/2)[ \ e^{+\lambda dt} +  e^{-\lambda dt} \ ]
= \cosh(\lambda dt) \ , 
$$
which gets bigger, indicating more phase volume, when averaged over the two possible
time directions.  In dynamical systems theory directions with growth are referred to
as the ``unstable'' manifold while the decay directions are the ``stable'' one.  If
the direction of $\delta$ is allowed to develop ``naturally'', with only its length,
but not its direction constrained, it soon comes to point in the direction associated
with the largest Lyapunov exponent $\lambda_1$ .

But this is not at all the direction of maximum growth.  To see this consider a simple
example\cite{b23}, a harmonic oscillator's orbit where the mass and the force constant are both
equal to 1/4, and with an orbit perturbation $(\delta q,\delta p)$ having a fixed length
$| \ \delta \ |$ .  If we choose a displacement parallel to the direction of the
perturbation ( as in Benettin's rescaling algorithm used to determine the maximum Lyapunov
exponent ) we need to solve two coupled evolution equations :
$$
\dot \delta_q = +4\delta_p - \lambda \delta_q \ ; \
\dot \delta_p = -(1/4)\delta_q - \lambda \delta_p \ \rightarrow
\lambda = (15/4)\delta_q\delta_p/| \ \delta \ |^2 \ ,
$$
subject to the fixed-length constraint
$\dot \delta_q\delta_q + \dot \delta_p\delta_p \equiv 0$ .
The maximum value, for equal perturbations in the two directions, gives a growth
rate of $(15/8)$ .  On the other hand the unperturbed growth rate in a {\it general}
direction, is $(4\delta_p,-(1/4)\delta_q)$ , which has its maximum value of $4$ for a
perturbation parallel to the $p$ direction.  {\it Maximum and minimum} growth rates
{\it are} time-reversible but the dependence of the {\it Lyapunov} growth and decay
rates are not. The latter rates depend upon past history, and not future destiny.

\subsection{Nonequilibrium Examples with Weak but Stiff Control of $p^2$ and $p^6$}

A specific ergodic system ,
$$
\dot q = p \ ; \ \dot p = -q - \zeta [ \ 1.5p - 0.5 (p^5/T^2) \ ] \ ;
$$
$$
\dot \zeta = 1.5[ \ (p^2/T) - 1 \ ] -0.5[ \ (p^6/T^3) - 5(p^4/T^2) \ ] \ , 
$$
exerts a passive control over the second and sixth velocity moments.  Passive in
that the equations are {\it necessarily consistent} with the Gaussian distribution
but are {\it not necessarily sufficient} for that distribution to be realized.
When the target temperature $T(q)$ varies with coordinate ,
$$
T(q) = 1 + \epsilon \tanh(q) \ ; \ 0 < \epsilon < 1 \longleftrightarrow 0 < T(q) < 2 \ ,
$$
the problem can become {\it dissipative}, with the comoving phase-volume rate-of-change
negative,
$$
\langle \ (\dot \otimes/\otimes) \ \rangle \equiv (\partial \dot q/\partial q) +
(\partial \dot p/\partial p) + (\partial \dot \zeta/\partial \zeta) =
\langle \ -1.5\zeta + 2.5\zeta (p^4/T^2) \ \rangle  < 0 \ .
$$
Sample cross-sections of the corresponding chaotic sea appear in {\bf Figure 10}.
At equilibrium, not shown in the Figure, a $(q,p,0)$ plot would show inversion
symmetry, equivalent to noting that viewing the thermostated oscillator in a
mirror would change the signs of $(q,p)$ without changing the friction coefficient
$\zeta$.  Changing the signs of the friction coefficient and both parameters
$(\pm \zeta,\pm 1.5,\mp 0.5)$ leaves the equations of motion unchanged.

\begin{figure}
\includegraphics[width=5in,angle=+90.]{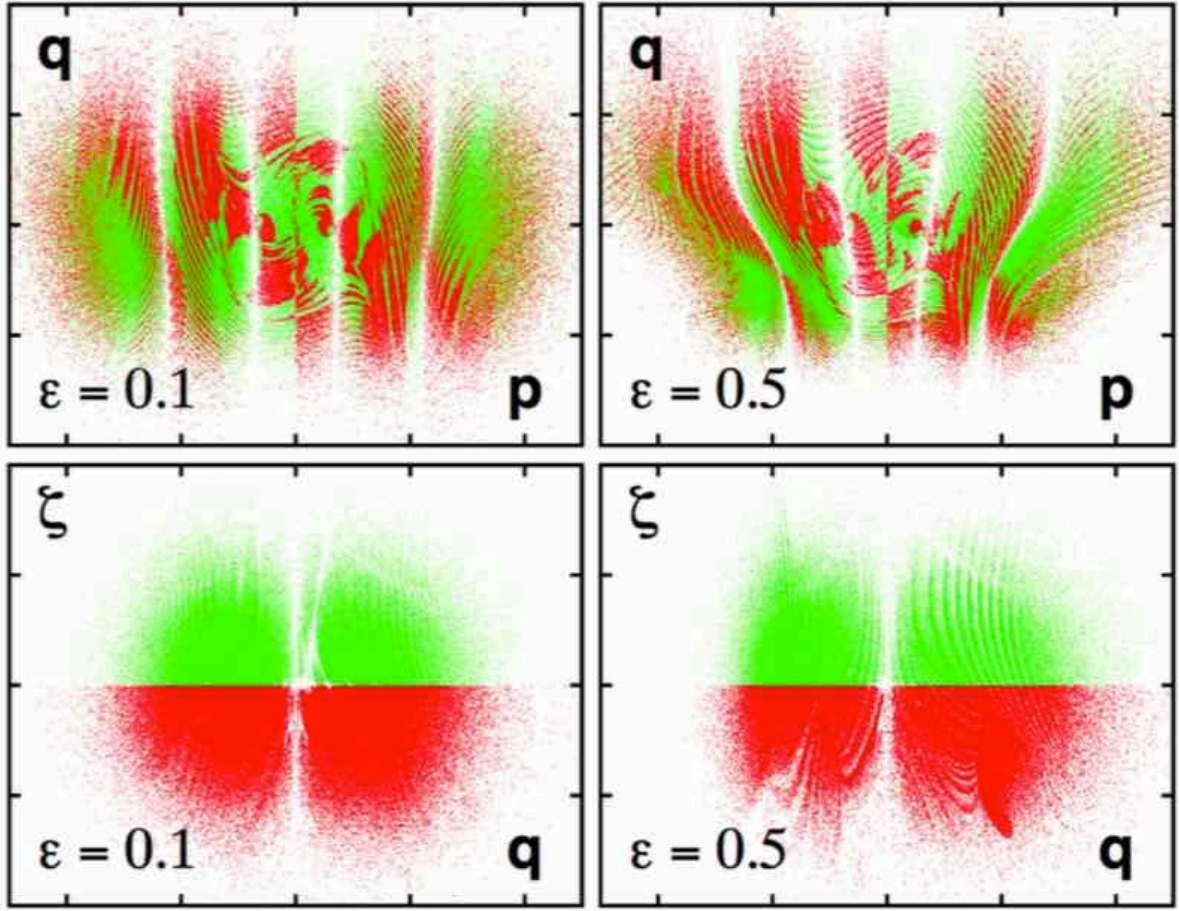}
\caption{
Cross Sections of the Dissipative Oscillator with $(\alpha,\gamma)=
(1.5,-0.5)$ . For all of the plots the abscissa ranges from  $-5.0 \
{\rm to} \ +5.0$ while the ordinate ranges from $-4.0 \ {\rm to} \
 +4.0$ .  The penetrations of the $\zeta = 0$ plane ( top panels )
corresponding to a positive largest Lyapunov exponent are in red, with
negative values in green.  The penetrations of the $p=0$ plane ( shown in the
lower panels ) are colored in the same way and confirm that negative values
of the friction coefficient correspond to phase-volume growth.
}
\end{figure}

An analytic calculation of the largest Lyapunov exponent for the $(\alpha,\gamma)$
problems illustrated in {\bf Figure 10} requires the solution of the three coupled
linearized equations of motion for the reference-to-satellite vector.  In the $p=0$
plane the equations for $\delta = ( \ \delta_q,\delta_p,\delta_\zeta \ )$ simplify :
$$
\{ \ \dot \delta_q = \delta_p - \lambda_1\delta q \ ; \
\dot \delta_p = - \delta q - 1.5\zeta\delta_p - \lambda_1\delta_p \ ; \
\dot \delta_\zeta = -\lambda_1 \delta_\zeta \ \} \ .
$$
The constraint of constant length imposed by $\lambda_1$ in its role as a
Lagrange multiplier, when applied in the $p=0$ plane, makes it possible to
relate the {\it sign} of $\lambda_1$ to that of the friction coefficient $\zeta$ :
$$
(d/dt)[ \ \delta_q^2 +\delta_p^2 +\delta_\zeta^2 \ ] \equiv 0 \longrightarrow
\lambda_1 = -\alpha \zeta\delta_p^2 \ .
$$
With $\alpha=0.05 \ {\rm or} \ 1.5$ the upper half plane of the lower panels of
{\bf Figures 10 and 11}, with $\zeta>0$ is entirely green, signifying a {\it negative}
Lyapunov exponent.  Positive values of $\lambda_1$ indicated by red in the Figures,
correspond to negative values of $\zeta$ in the $p=0$ plane.  As would be expected
for a ``friction'' coefficient negative values promote growth and positive ones decay.
As the temperature gradient increases the dissipation grows.  Because the
time-averaged dissipation is necessarily positive a comoving volume element
$dqdpd\zeta$ vanishes with time, exponentially fast.

\subsection{Nonequilibrium Example with Weak Control of $p^2$ and $p^4$}

Because the equations of motion for control of the sixth moment are stiff,
requiring a Runge-Kutta timestep of order 0.0001 or 0.0002, we consider the
simpler case of passive control of the second and fourth velocity moments :
$$                         
\dot q = p \ ; \ \dot p = -q - \zeta [ \ 0.05p + 0.32(p^3/T) \ ] \; 
$$
$$
\dot \zeta = +0.05[ \ (p^2/T) - 1 \ ] + 0.32[ \ (p^4/T^2) - 3(p^2/T) \ ] \ .
$$
Like those controlling $\langle \ p^2,p^6 \ \rangle$ these equations
controlling  $\langle \ p^2,p^4 \ \rangle$
are ergodic at equilibrium, and are consistent with Gibbs' distribution
$f(q,p,\zeta) \propto e^{-q^2/2T}e^{-p^2/2T}e^{-\zeta^2/2}$ . 

Away from equilibrium, with $T = 1 + \epsilon \tanh(q)$ and where tori are
absent, these same equations generate {\it dissipative} fractal attractors.
For small temperature gradients the dimensions of the cross sections in
{\bf Figure 11} and the dissipation vary smoothly while the motion remains
ergodic.  Just as in the Galton Board examples the resulting fractals signal
the rarity of nonequilibrium states.

Because the two-parameter motion equations {\it are} time-reversible there exists
a symmetric set of $\zeta = 0$ cross-section states with reversed momenta $\pm p
\longleftrightarrow \mp p$ , reversed heat current, reversed dissipation, as well
as phase-volume growth rather than collapse, all of these characteristics 
violating the Second Law of Thermodynamics.  But all those states are unstable,
with a reversed ( when time-averaged ) Lyapunov spectrum having a positive
sum.  They make up the unobservable and illegal {\it repellor states}.  The repellor
acts as a source and the attractor a sink for these time-reversible heat-flow
problems.  Despite their symmetry the repellor and attractor measures are different,
zero and one, respectively.

\begin{figure}
\includegraphics[width=5in,angle=+90.]{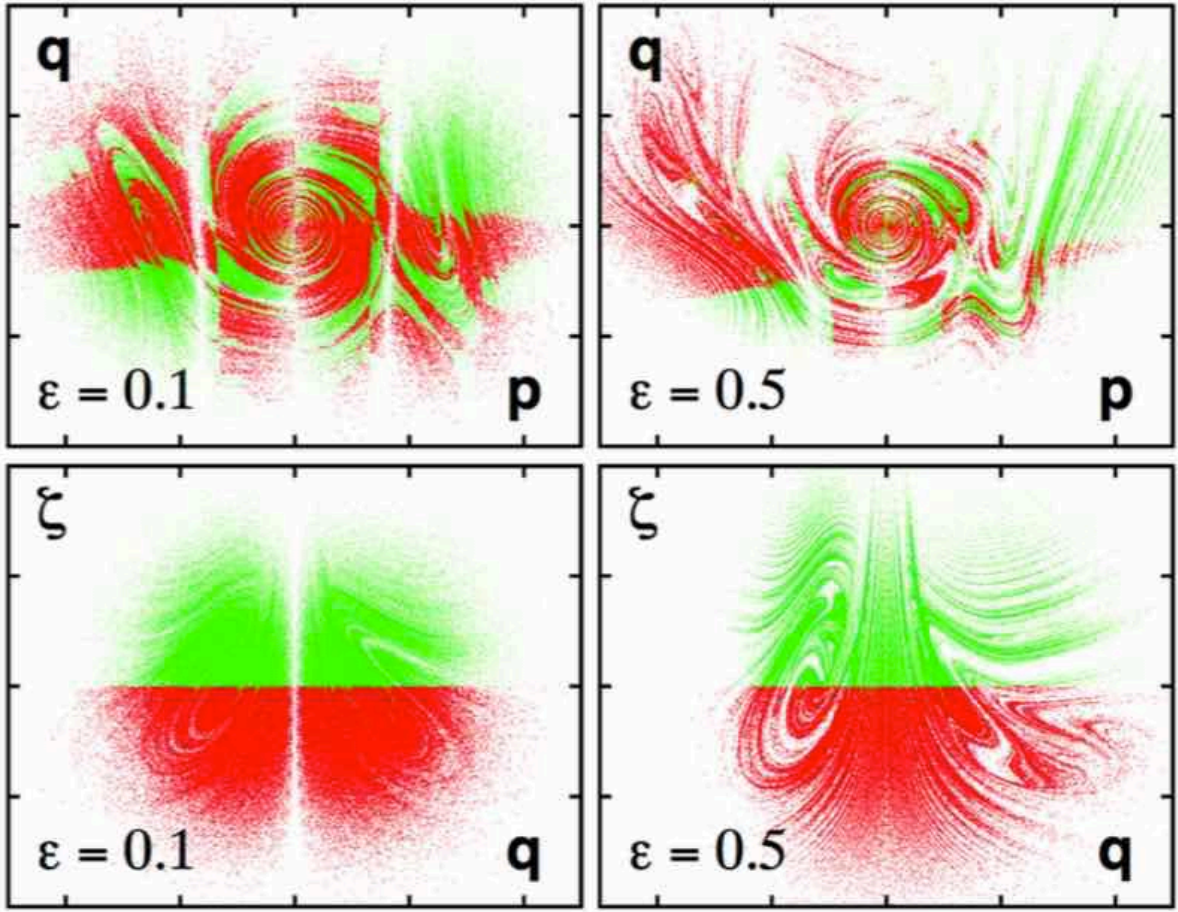}
\caption{
Cross Sections of the Dissipative Oscillator with $(\alpha,\beta)=
(0.05,0.32)$ and a friction linear in $\zeta$ $(n=1)$ .  For all of the plots the abscissa ranges from  $-5.0 \
{\rm to} \ +5.0$ while the ordinate ranges from $-4.0 \ {\rm to} \  +4.0$ . 
}
\end{figure}

\begin{figure}
\includegraphics[width=5.5in,angle=-90.]{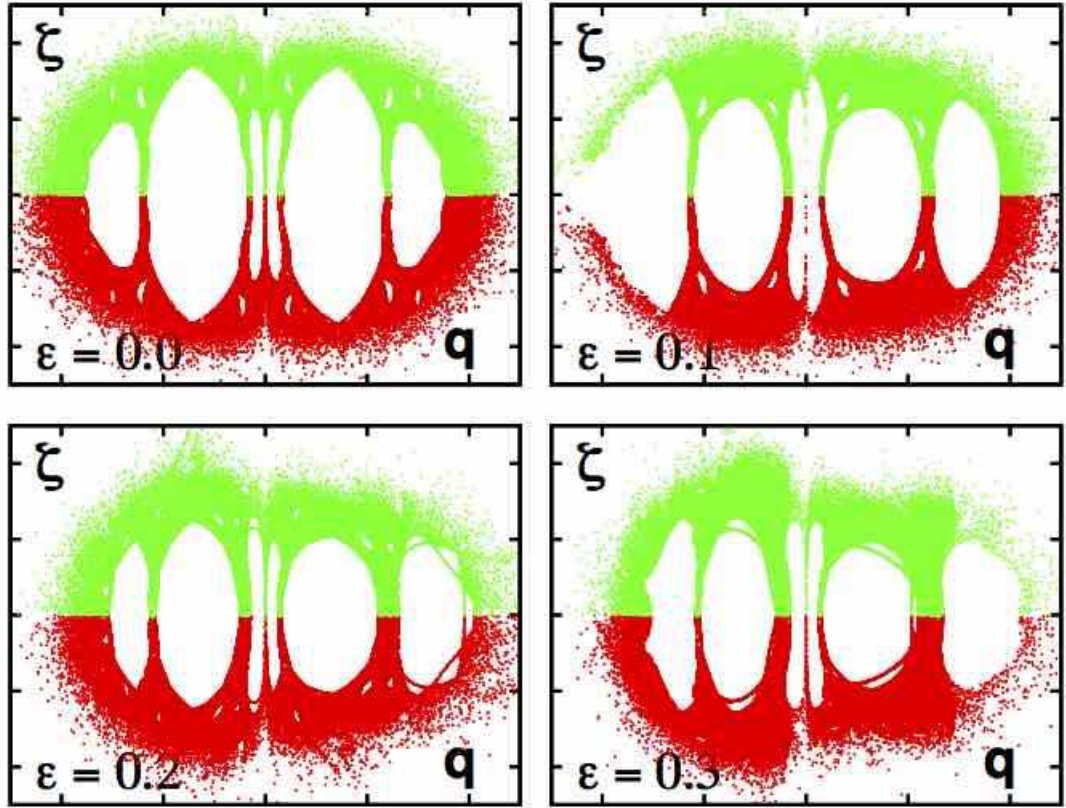}
\caption{
The effect of dissipation on phase-space cross sections for the chaotic, but
nonergodic, Nos\'e-Hoover oscillator.  Sections with vanishing momentum are
shown for four values of $\epsilon$ : $0.00, \ 0.10, \ 0.20, \ 0.30$ . The
equations of motion are $\dot q = p$ ; $\dot p = - q - \zeta p$ ; $\dot \zeta
= [ \ (p^2/T) - 1 \ ]$ , where $T = 1 + \epsilon \tanh(q)$ . 
}
\end{figure}

The implication of Figure 11 is relatively simple.  Starting out with an
equilibrium dynamics which follows Gibbs' canonical distribution ( without
holes ) relatively simple multifractal attractors respond to a thermal
gradient.  Increasing the temperature gradient leads to a reduced attractor
dimensionality and increased dissipation.  The general approach of perturbing
an ergodic equilibrium Gibbs' ensemble evidently leads to relatively simple
nonequilibrium steady states.  In an effort to see whether or not this simplicity
has a counterpart in nonergodic Hamiltonian mechanics we return to the
Nos\'e-Hoover oscillator and expose it to a thermal gradient next.

\subsection{Nonequilibrium Nos\'e-Hoover Oscillator Dynamics}

Although the equilibrium Nos\'e-Hoover oscillator has {\it all} the complexity of
Hamiltonian mechanics -- chains of islands and their elaborations, it is
worth exploring whether or not any simplification results away from equilibrium.
Our previous work\cite{b25} indicated a crude boundary between small-gradient
and large-gradient behavior around $\epsilon = 0.40$ .  In {\bf Figure 12} we
consider strange-attractor solutions corresponding to three values of
$\epsilon, \ 0.10, \ 0.20, \ 0.30$ in addition to the equilibrium case.  We
use nonequilibrium versions of the Nos\'e-Hoover oscillator with $\dot \zeta =
[ \ (p^2/T) - 1 \ ]$ rather than $\dot \zeta = [ \ p^2 - T \ ]$ because the
multimoment models are {\it much} simpler to formulate when the distribution of
the friction coefficient(s) is not explicitly temperature dependent.

We considered relatively long runs ( $10^{12}$ timesteps ) in an effort to
assure convergence.  A bit more than halfway through the $\epsilon = 0.10$
simulation, with a fourth-order Runge-Kutta timestep of 0.01, the chaotic
strange attractor suddenly began to generate a torus which then gradually shrunk
with time.  Was this real, or an artefact? A check calculation with fifth-order
Runge-Kutta, also using a timestep of 0.01 revealed no such behavior, exhibiting
instead a chaotic solution. This problem illustrates the virtue of comparing
results from the two or more integrators, particularly when longer runs are
desirable.  A careful investigation shows that the single-step fourth-order and
fifth-order errors:
$$
{\rm RK4 \ error} \ \simeq -dt^5/120 \ ; \
{\rm RK5 \ error} \ \simeq +dt^6/720 \ ,
$$
are in opposite directions, with the fifth-order still noticeable with a
timestep $dt = 0.01$ in double-precision calculations.  With $dt = 0.001$
both the fourth-order and fifth-order  errors are negligible in
double-precision work.  For the reader interested in exploring these small
effects\cite{b26} an initial condition very close to the border between chaos and
tori is $(q,p,\zeta) = (\delta,\delta,3)$ with $\delta$ small.  A small
nonzero value of $\delta$ ( such as $10^{-12}$ ) is necessary to avoid the
analog of a $(q,p)$ fixed point in the $(q,p,\zeta = 0)$ cross section.

{\bf Figure 12} shows the sign of the local Lyapunov exponent $\lambda_1(t)$
in color, both at and away from equilibrium ( $\epsilon = 0.0$ to 0.3 ).
Notice that the near inversion symmetry in the $(q,p,0)$ plane for
$\epsilon = 0.10$ , gives way to predominating Lyapunov instability far from
equilibrium, at $\epsilon = 0.30$ .  Below, the $(q,0,\zeta)$ plane shows
that the sign of the local Lyapunov exponent is a perfect match of the sign of
the local friction coefficient.  Simply reversing the direction of the flow
in the three nonequilibrium panels,
corresponding to reflection of $p$ about the $q$ axis, might be expected to
change the signs of the Lyapunov exponents, but even close to equilibrium
this does not happen.  This is because that exponent depends upon the {\it past}
so that there is a fundamental lack of symmetry in the local exponents.

\section{Liouville's Theorem Applied to Nonequilibium Flows}

We have seen that the continuity equation is an invaluable tool in finding
constrained dynamical systems consistent with Gibbs' canonical ensemble. This
idea was used by Green and Kubo to express transport coefficients in terms of
equilibrium fluctuations.  Nonequilibrium simulations, even far from
equilibrium, use this same tool.  Because the Galton Board as well as all of the
thermostated oscillator problems we have considered involve
{\it three}-dimensional flows it is natural to consider their analysis and display
from the standpoint of Liouville's phase-space continuity equation :
$$
(\partial f/\partial t) = -\nabla \cdot (fv) \longleftrightarrow
\dot f = (\partial f/\partial t) + v\cdot \nabla f \equiv - f\cdot \nabla v \ ,
$$
also in three dimensions.  The corresponding motion equations are represented by
$v \equiv (\dot q,\dot p, \dot \zeta)$ in the oscillator problems.

It is tempting to imagine solving the flow equation directly, replacing the
derivatives $(\dot q,\dot p,\dot \zeta)$ by finite differences.  Our colleague
John Ramshaw made us the welcome present of his ``upwind-differencing'' computer
program, which transfers density across all six faces of each cubic computational
cell according to the velocities evaluated at the cell boundaries, determining which
of any two adjacent cells is the ``donor'' of probability, and which is the 
``acceptor''.  The flow is taken to be proportional to the {\it donor} probability
though it would appear that an average probability is nearly as plausible.  Using
the average, however, leads to exponential instability\cite{b27}.  {\it At} equilibrium,
or in a steady state, the six flows into and out of every cell must balance.
Evidently the algorithm conserves probability exactly, but {\it not} time-reversibly.
Liouville's continuity equation {\it is} time-reversible.  But it is easy to see that
in a reversed implementation of the flow algorithm the cells furnishing the
probability forward in time will not have it returned exactly in the ``reversed''
step.

The time-reversal thought experiment ( as well as its computational realization )
show that the algorithm has a diffusive irreversibility.  To see this begin with
a single filled zone.  Going forward in time this zone would typically donate 
probability to half (three)  of its neighbor zones.  The other three neighbors
would remain empty.  On reversal most of the probability gained by the three
neighbors would remain there.  Such a forward-plus-backward two-timestep simulation
reveals why it is that a straightforward application of Liouville's flow equation
fails.

A simple computational test of this Liouville algorithm is to reproduce the
periodic conservative Nos\'e-Hoover orbit, shown in {\bf Figure 5} . The initial
conditions $(q,p,\zeta) = (0.00,1.55,0.00)$ generate it.  Solving the
Liouville equation with {\it all} of the initial density in the cells
nearest the initial condition leads to an amplitude loss proportional to
the cell size. After a few such damped oscillations the final quiescent
stationary state is obtained, with most of the density bunched near the
origin. {\bf Figure 13} displays the mean values of the finite-difference
algorithm's history for ten oscillation periods.  Limit-cycle and chaotic
problems lead to very similar results.

\begin{figure}
\includegraphics[width=4.0in,angle=+90.]{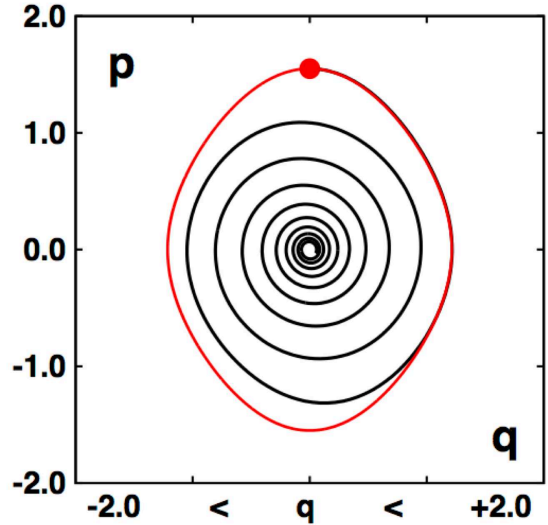}
\caption{
The Nos\'e-Hoover periodic orbit  with $T = 1$ , as in {\bf Figure 5} , is
shown in red, with a bullet indicating the initial condition, $(q,p,\zeta) =
(0.00,1.55,0.00)$ .  The finite-difference Liouville
continuity equation solution for
$\langle \ p(\langle \ q \ \rangle ) \ \rangle$
is shown for a time of 55.78, ten oscillation periods, in black, using a
$40\times40\times40$ mesh with a mesh spacing of $0.1$ . Solutions with three
finer meshes ( 0.0500, \ 0.0250, \ 0.0125 ) establish that the artificial damping is
proportional to the mesh size as is shown in John Ramshaw's book, {\it Elements
of Computational Fluid Dynamics}\cite{b27}.
}
\end{figure}

A better rendition of the continuum motion equations could follow a set
of Lagrangian points, attracted toward one another with a Lagrange
multiplier designed to follow the comoving flow volume precisely.  Similar
algorithms, based on interface tracking could perhaps be developed, but in
the end the numerical solution of Liouville's continuity equation appears
much better suited to developing consistent motion equations than to
evolving a continuous phase-space density.

\section{Analogies with Many-Body Problems}

Over the years thousands of papers have described the use of time-reversible
control variables to solve simple problems. Shear flows, heat flows, and
shockwaves are examples that spring to our minds.  Large-scale biomolecular
simulations use exactly these same ideas.  In stationary nonequilibrium flows
 the atomistic forces are supplemented by boundary, constraint, and driving
forces in such a way as to generate a nonequilibrium steady state.  In every
case the resulting phase-space distribution is fractal, representing a flow
from a strange repellor, with a positive summed Lyapunov spectrum to a
strange attractor, with a negative Lyapunov sum.  Although the fundamentals
are no more complicated than the examples detailed in this paper, the
possibilities for more complex applications are and will continue unlimited.

\section{Acknowledgments}
We wish to thank Puneet Patra and Baidurya Bhattacharya for their fresh look
at ergodicity described in their 2014 and 2015 Journal of Chemical Physics
papers\cite{b28,b29}.  This stimulation led to some of the advances described
here.  We also thank John Ramshaw for useful discussions relating to
Liouville's Theorem as well for furnishing us a computer program providing the
finite-difference results described in this paper.

\end{document}